\documentclass[preprint,aps]{revtex4}
\usepackage{graphicx}%

\begin{document}
\title{Probability of stochastic processes and spacetime geometry}
\author{Enrique Canessa\footnote{e-mail: canessae@ictp.it}}
\affiliation{The Abdus Salam International Centre for Theoretical Physics,
Trieste, Italy
\vspace{3cm}
}

\begin{abstract}
We made a first attempt to associate a probabilistic description 
of stochastic processes like birth-death processes
with spacetime geometry in the Schwarzschild metrics on distance 
scales from the macro- to the micro-domains.  
We idealize an {\it ergodic} system in which system states
communicate through a curved path composed of transition arrows 
where each arrow corresponds to a positive, analogous 
birth or death rate.
\\
\\
To appear Physica A (2007)
\end{abstract}
\maketitle

It is well known that stochastic processes can occur according to 
a simple 1D {\it birth-and-death} model (see, {\it e.g.}, 
\cite{Goo88,Nis82}).
The probabilities $p_{j}$ at the system state $j$ in this model are
obtained recursively from the relation
\begin{equation}\label{eq:bd2}
{\hat \mu}_{j} \; p_{j}(t) = 
    {\hat \lambda}_{j-1} \; p_{j-1}(t) \;\;\; (j=1,2\cdots) \;\;\; ,
\end{equation}
where by a choice of the birth ${\hat \lambda}_{j}$ and death 
${\hat \mu}_{j}$ rates various stochastic models can be 
constructed, {\it e.g.}, cell populations, queueing, {\it etc}.
Birth-death processes are Markow stochastic processes where the states 
$1 \le j \le n$ represent the size of a fluctuating population 
(of system capacity $n$ or greater)
limited to births and deaths with $p_{j}(t)$ population probability 
distributions.  
The infinitesimal population variation $dn$ occuring during a time
interval $dt$ is a {\it random variable} with possible
values $\pm 1$.  Its statistical expectation value is
$E\{dn\} = (+1) \times \{Prob. \; of \; a \; birth \; at \; dt\} \; + \;
(-1) \times \{Prob.\; of\; a \; death \; at \; dt\}$ , which implies
\begin{equation}\label{eq:bd3}
E\{dn\} = ({\hat \lambda}_{n} - {\hat \mu}_{n}) \; dt\;\;\; .
\end{equation}
The actual change in population of any one replicate can be written as 
$dN/dt = {\hat \lambda}_{n}-{\hat \mu}_{n} + \eta(t) ({\hat \lambda}_{n}-{\hat \mu}_{n})^{1/2}$ where $\eta(t)$ is a random variable with zero mean
and  unit variance.  Thus the expected system capacity change 
during $dt$ is the same as would be predicted by a determinitic model.

In this work we shall associate the probabilistic description of 
Eq.(\ref{eq:bd2}) of processes describing the behaviour of a stochastic
physics system to space curvature within the Schwarzschild description.  
The idea that the stochastic birth and death rates can be associated
to some geometric properties of spacetime derivated from general
relativity framework is quite new and interesting.
We believe this association could be useful 
to imagine alternative views of spacetime in a more comprehensible 
way, even though the notion of probability seems to be purely 
epistemic within a theory of spacetime geometry like 
relativity \cite{Sau01}. 

The infinitesimal radial distances in the Schwarzschild metrics,
at fixed polar angles $\theta$ and $\phi$ and time $t$,
satisfy \cite{Fos94}
\begin{equation}\label{eq:dr}
dR \equiv (1-2m/r)^{-1/2} dr \;\;\; ,
\end{equation}
where $m \equiv GM/c^{2}$, and $M$ is the mass of the body producing
the field, $G$ is the Gravitational constant and $c$ is the speed of
light.  The Schwarzschild isotropic solution is the basics for the
tests of general relativity, including the bending of the light and
the spectral-shift.

This spacetime metric which 
leads to an exact solution of the Einstein field equations of 
general relativity in a static, spherically symmetric gravitational 
field has been assumed to be valid even at microscopic scales.  
In particular, the superposition of ($N$-electrons) quantum states over 
the $10 cm$ scale human brain has been estimated to last for millions 
of years using a free electron Schwarzschild radius of about 
$1.35\times 10^{-55} cm$ \cite{Pen89}.  On the 
other hand, elementary particles have also been studied under the 
perspectives of microscopic black holes \cite{Bek98,Mat04}. 
This microworld scheme suggests that the black hole mass spectrum 
in quantum theory is discrete such that the black hole horizon area 
is confined to equaly spaced levels whose degeneracy corresponds to
the black hole entropy \cite{Hak88}.  

Thus, in principle, the universe we live in may consist of 
an inhomogeneous density of matter (in the form of starts, molecules, 
atoms, {\it etc}), that curve space due to their gravitational fields 
on scales ranging from the size of the observable universe
to the micro-world. 
Intuitively, all of this makes sense at least up to a submolecular
order of magnitude in which the force of gravity cannot be negligible
compared to other fundamental forces found in nature.

It is not unreasonable to consider then the possibility of 
extending the Schwarzschild metric to the vicinity of multiple 
massive objects, due to the inhomogeneous presence of matter in 
the universe at different geometrical scales. 
When $M \ne 0$, the curvature of space implies that $dR > dr$ 
and the coordinates not necessarily have the 
physical meaning as they have in a flat space where time 
is measured by clocks -stationary in the reference system employed, 
and $r$ is a radial distance from the origin. 

As shown in Fig.1 (for each of the interconnected surfaces 
illustrated), $S_{o}$ represents a portion of flat space 
(with $M \longrightarrow 0$), while the (nested) curved surface $S_{m}$ 
represents a portion of the curved space.  The circles $C_{1}$ and 
$C_{3}$ represent spheres of radius $r$ while the circles $C_{2}$ 
and $C_{4}$ represent spheres of radius $r+dr$, embedded in an 
Euclidean space. 

Let us extend Eq.(\ref{eq:dr}) and
consider a system of nested curved surfaces $S_{m}$ to form 
a spiral of $n$-curved surfaces as illustrated in the figure which 
are interconnected in the sense that 
\begin{eqnarray}\label{eq:dr1}
dR^{(1)} & \equiv & (1-2m^{(1)}/R)^{-1/2} dR \;\;\; , \nonumber \\
dR^{(2)} & \equiv & (1-2m^{(2)}/R^{(1)})^{-1/2} dR^{(1)} \;\;\; , \nonumber \\
\cdots  \;\;  &  & \;\; \cdots   \nonumber \\
dR^{(n)} & \equiv & (1-2m^{(n)}/R^{(n-1)})^{-1/2} dR^{(n-1)} \;\;\; , 
\end{eqnarray}
In other words we are simply assuming the space to be curved 
at all geometric scales $dR^{(n)}$ due to the presence of 
scattered matter $M^{(i)} \ne 0 \;(i=0,1,\cdots,n$) with 
$M^{(0)}\equiv m$.  It is enough to select an infinitesimal 
$dR^{(n)}$ for large $n$ in order all other radial distances 
$dR^{(n)} > \cdots > dR > dr$ to be also infinitesimal.

Let us define next the $R$-dependent trial functions $\lambda$ 
and $\mu$, to be indentified below, as
\begin{eqnarray}\label{eq:lambdamu}
\lambda^{(n)}  & \rightleftharpoons & ( R^{(n)} - R^{(n+1)}_{s} )/r_{s}  \nonumber \\
\mu^{(n)}        & \rightleftharpoons &  R^{(n-1)}/r_{s} \;\;\; , 
\end{eqnarray}
where $R^{(n)}_{s}$ are Schwarzschild radii at which the metric 
of Eq.(\ref{eq:dr}) becomes singular for positive $n \ge 1$, namely
\begin{equation}\label{eq:rs}
R^{(n)}_{s} = { 2GM^{(n)} \over c^{2} } \;\;\; , 
\end{equation}
such that $R^{(0)}_{s} \equiv r_{s} = 2m$.  The ratio 
$\lambda/\mu$ tends to $0$ as 
$R \rightarrow R_{s}$ and it tends to unity as $R \rightarrow \infty$ 
(or, alternatively, $M^{(n)} \rightarrow 0$).  Using 
Eqs.(\ref{eq:dr1}) and (\ref{eq:lambdamu}), the ratio of their 
product satisfies in particular
\begin{eqnarray}\label{eq:p0p1}
{ \lambda^{(1)}\lambda^{(0)}\over \mu^{(2)}\mu^{(1)} } \; {\cal P}^{(0)}(r) & \rightleftharpoons &
        \left( { R^{(1)} - R^{(2)}_{s} \over R^{(1)} } \right)
        \left( { R - R^{(1)}_{s} \over R } \right)
   \left( { r - r_{s} \over r } \right)  \nonumber \\
  & = &
        \left( { dR^{(2)} \over dR^{(1)} } \right)^{-2}
        \left( { dR^{(1)} \over dR } \right)^{-2}
   \left( { dR \over dr } \right)^{-2} \nonumber \\
  & = &
        \left( { dr \over dR^{(2)} } \right)^{2} \;\;\; ,
\end{eqnarray}
where we have set $R^{(0)} \equiv R$ and
${\cal P}^{(0)}(r)  \equiv (r - r_{s})/r \equiv (dr/dR)^{2} < 1$. 

It is then straightforward to establish the general recursive relation
\begin{equation}\label{eq:bd1}
{\cal P}^{(n)}(r) \rightleftharpoons 
{ \lambda^{(n-1)}\cdots \lambda^{(1)} \lambda^{(0)} \over \mu^{(n)}\cdots \mu^{(2)}\mu^{(1)} }  
\; {\cal P}^{(0)}(r) \;\;\; ,
\end{equation}
such that
\begin{equation}\label{eq:bd}
\mu^{(n+1)} \; {\cal P}^{(n+1)}(r) = \lambda^{(n)} \; {\cal P}^{(n)}(r) \;\;\; ,
\end{equation}
and
\begin{equation}\label{eq:master}
\left( { dr \over dR^{(n)} } \right)^{2} \equiv 
           {\cal P}^{(n)}_{\theta,\phi,t}(r) < 1 \;\;\; ,
\end{equation}
at fixed time and (along a fixed path of) polar angles.  
The latter is our master equation 
relating space geometry to an iterative ($n$-process) function.

In view of the above we could then 
associate Eq.(\ref{eq:bd}) to Eq.(\ref{eq:bd2}) in analogy to birth-death 
types of processes (projection-like operators).  The analogous 
'birth' and 'death' rates being defined by Eq.(\ref{eq:lambdamu}) 
when the system is in a given state $n$.
The left side of Eq.(\ref{eq:bd}) becomes an analogous effective 
rate at which the system
is leaving the state $n$ because of the analogous decaying
processes (deaths), and the right side becomes the effective rate 
at which the system is entering state $n$ because of analogous births.

Within this analogy, a large number of aggregating processes
would change as a function of distance $R$ acting collectively 
to pass on information or to introduce system disorder   
(via the information like entropy 
$-{\cal P}^{(n)}\ln {\cal P}^{(n)} >0$).
In this case from the definition of the Schwarzschild radii of 
Eq.(\ref{eq:rs}) we write for the scattered matter
\begin{equation}\label{eq:mass}
 { M^{(n+1)} - M^{(n)} \over M } =  { R^{(n+1)}_{s} - R^{(n)}_{s} \over r_{s} }  
 \rightleftharpoons  (\mu^{(n+1)} - \mu^{(n)}) - (\lambda^{(n)} -\lambda^{(n-1)}) \;\;\; , 
\end{equation}
or, by using ${\cal E}^{(0)} \equiv {\cal E} \rightarrow Mc^{2}$ and
forward and backward two-point derivative approximations,
\begin{equation}\label{eq:energy}
 \frac{1}{{\cal E}}\left(\frac{d{\cal E}^{(n)}}{dn}\right) \approx \frac{d\mu^{(n)}}{dn} - \frac{d\lambda^{(n)}}{dn}\;\;\; .
\end{equation}
Since $M$ is finite, then $\mu^{(n+1)} \ne \lambda^{(n)}$ or
$\mu^{(n)} \ne \lambda^{(n-1)}$. 
In a flat space $dR \longrightarrow dr$, then system 
is in an analogous state $n=1$ and $\lambda \rightarrow \mu$.

The above relation means that variations of an analogous coefficient 
of system energy could be related to the difference between some 
'annihilated' analogous processes of leaving state $n$ and those being 
'generated' in the state $n$.  If this is the case, we may look at 
the function ${\cal P}^{(n)}$ in Eq.(\ref{eq:master}) as an analogous 
probability measure provided the sum over $n$ different states 
satisfies the normalization condition
\begin{equation}\label{eq:master1}
\sum_{i=0}^{n} {\cal P}^{(i)}(r) = \sum_{i=0}^{n} \left( { dr \over dR^{(i)} } \right)^{2} = 1 \;\;\; .
\end{equation}
Let us analyse next if this normalization condition is rigorously valid.
From Eq.(\ref{eq:bd1}) we can see that ${\cal P}^{(n)}$ is completely 
determined by the value of ${\cal P}^{(0)}$ at each distance.  
On physics grounds we have that $r > r_{s}$, hence 
$1 > {\cal P}^{(0)} \equiv (r - r_{s})/r  > 0$ and
${\cal P}^{(n)} > 0$.  This means that 
the normalization condition for an anologous topological probability 
directly relates the singularity of Schwarzschild's metrics 
(black hole horizon) $r_{s}$.  Then it only remains to choose 
${\cal P}^{(0)}$, such that the associated (or analogous) probabilities 
$\{ {\cal P}^{(i)} \}$ sum to unity $\forall r$.  Similarly to 
birth-death stochastic processes \cite{Goo88,Nis82}, for large $n$ 
one can choose 
\begin{equation}\label{eq:rho}
{\cal P}^{(0)}(r) \equiv \left( { dr \over dR } \right)^{2}  =
  \left( 1 + \sum_{i=1}^{n} \rho_{i} \right)^{-1} < 1 \;\;\; , 
\end{equation}
where $\rho_{i}$ is defined by $i$-products of birth-death ratios 
$\lambda^{(i-1)}\cdots \lambda^{(0)} / \mu^{(i)}\cdots \mu^{(1)}$.  
For $n \ge 1$, these ratios get small sufficiently rapidly because 
the trial associations in Eq.(\ref{eq:lambdamu}) plus the fact that
$R > 0$ $\forall i$, imply the limits
$ 0< \lambda^{(i-1)}/\mu^{(i)} = 1 - (R^{(i)}_{s}/R^{(i-1)}) < 1$.  
Hence $\sum_{i=1}^{n} \rho_{i} < \infty$ and
$\sum_{i=0}^{n} {\cal P}^{(i)}(r) \equiv \sum_{i=0}^{n} \rho_{i} {\cal P}^{(0)}(r) = [1 + \sum_{i=1}^{n} \rho_{i}] \dot  {\cal P}^{(0)}(r) = 1$.
This also implies a well-defined stationary state to exist and
that the general sequence defined by Eq.(\ref{eq:bd}) 
is the unique normalized solution of an equivalent equation of motion
of a population probability distribution \cite{Goo88,Nis82}.  

We can also derive a recurrence relation for the universal 
time coordinate.  Infinitesimal proper time intervals
for a clock at a fixed distance in space are given by 
$d\tau \equiv (1-2m/r)^{1/2} dt$, such that in the curved space
$d\tau < dt$.  By a similar analysis to that leading to 
Eq.(\ref{eq:bd}), we obtain $(d\tau^{(n)}/dt)^{2} < 1$
after $n$-iteration processes. In terms of a temporal translation
this quantity is always positive at different scales. This could 
be seen as a forward positive direction for the "real" time in 
the direction in which disorder ({\it i.e.}, positive entropy) 
increases.

The trial functions defined by Eq.(\ref{eq:lambdamu}) are strictly
deterministic.  Next, they have been turned stochastic via 
Eq.(\ref{eq:bd}).  So what is the source of randomness in
our approach?  The answer is as follows.  Integrating 
Eq.(\ref{eq:energy}) with respect to any one of a countable number 
of system states $n$, and using Eq.(\ref{eq:lambdamu}), it gives
\begin{equation}\label{eq:energy2}
 { {\cal E}^{(n)} \over {\cal E} } + cte \approx 
       \mu^{(n)} - \lambda^{(n)} \rightleftharpoons
  { R^{(n-1)} - R^{(n)} + R^{(n+1)}_{s} \over r_{s} } \approx
 - \frac{1}{r_{s}}\left(\frac{dR^{(n)}}{dn}\right) 
  + { R^{(n+1)}_{s} \over r_{s} } \;\;\; .
\end{equation}
By comparison with Eq.(\ref{eq:bd3}) we then readily identify
\begin{eqnarray}\label{eq:source}
\frac{dN}{dt}  & \rightleftharpoons & \frac{1}{r_{s}}\left(\frac{dR^{(n)}}{dn}\right) \;\;\; , \nonumber \\
\eta(n) \left(\lambda^{(n)} - \mu^{(n)}\right)^{1/2}  & \equiv & 
      { R^{(n+1)}_{s} \over r_{s} } \;\;\; .
\end{eqnarray}
This means that the infinitesimal geometrical variation $dR^{(n)}$ 
occuring during a system state interval $dn$ corresponds to an 
analogous {\it random variable} and some 
$0< \eta < 1$ fluctuations in the Schwarzschild radii $R^{(n)}_{s}$
are our source of randomness.  Generally randomness
for {\it birth-and-death} processes can come from 
competition selection \cite{Sta05}, environment pressure \cite{Aus04},
random particle fluctuations \cite{Men06} or random noise \cite{Doe04}.

Our goal here has been to connect some concepts from stochastic processes
(in the form of birth/death rates) to concepts of topology, in
particular to space curvature in the Schwarzschild simplest 
(isotropic) metrics.  Thus, the starting points
have been the birth-death generic equation, Eq.(\ref{eq:bd2}) and the 
Schwarzschild metrics, Eq.(\ref{eq:dr}).  The core of the paper
is the general recursive relation for Schwarzschild radii,
Eq.(\ref{eq:bd}) in which a probability is assigned to the first
term of the recurrence and two trial functions geometrically
defined in Eq.(\ref{eq:lambdamu}), are assimilated to the birth
and death rates.  In this way, Eq.(\ref{eq:bd}) becomes similar
with a {\it birth-and-death} stochastic equation, and allows for
defining the system statistical energy (and thereof the entropy) of 
Eq.(\ref{eq:energy}) such that the normalization condition of 
Eq.(\ref{eq:master1}) implies to deal with distance scales greater 
than the Schwarzschild radius $r_{s}$. The ratios 
$\lambda^{(n-1)}/ \mu^{(n)}$ decay to zero sufficiently rapidly 
for $n \ge 1$.  We have 
assumed that the system is {\it ergodic} --{\it i.e.}, all $n$-states
of the system communicate through a curved path composed of 
transition arrows, each arrow corresponding to a positive birth or 
death rate.

In the light of this simple description different 
physics theories --corresponding to different averaging scales 
(from, say, particle physics to astrophysics),
may be associated via a geometric approach to 
probabilities \cite{Ell05}. One could deduce the 
metric, which allows a non-inertial observer to perform experiments 
in spacetime \cite{Min05}, via a probabilistic approach 
and vice versa.  However, to understand the underlying order 
in the world a metric independence would be more encopassing.

We regard the present ideas as a toy approach only. 
Toy models to search for a solution to spacetime problems 
in physics are not new.  An example is the toy model 
providing  an  extension  of the dimensionality of spacetime,
with an additional spatial dimension which is macroscopically
unobservable \cite{Tar01}. 

So far our approach has no output variables.  It could become more
reliable if some universality of generation-annihilation mechanisms 
are found to have similar features at the microscopic, mesoscopic 
and microsopic scales as suggested by Fig.1.  An indication of 
universality though is given by the general fluctuation 
theorem derived from Jarzynski's inequality for univariate 
birth-death processes driven out of equilibium by the external 
variation of a control parameter \cite{Sei04,Jar05}.  This is so 
because the theory of {\it birth-and-death} processes has been applied 
independently on astrophysics \cite{Bea06}, biophysics \cite{Nov06} 
and quantum physics \cite{Gas99} and the fluctuation theorem
becomes valid in each of these situations.

The recurrence equations (\ref{eq:bd}) can be interpreted as
an equivalent {\it conservation of radial distance flow} relation
({\it i.e.}, rate up = rate down).
That is, the long radio rate at which the system of nested surfaces
moves up from state $n$ to $n+1$ equals the rate at which the system
moves down from state $n+1$ to $n$. Thus our analogous birth-death
process describes an evolution in radial distance whose values 
increases or decreases (stochastically) by one single state starting 
from an analogous ground probability state ${\cal P}^{(0)}$.
In an universe seen at different 
scales, averaged effects of small-scale inhomogenities may alter both
observational and dynamical relations at the larger scale
\cite{Ell05a}.  Probability could be then consistent with the full 
filling of the whole space with nested curved surfaces.  There is
no a priori specified time and one would ultimately expect measured 
time to be increasingly positive or future oriented.

It is tempting to relate this first attempt, based on gravity and 
stochastic processes associated via Eq.(\ref{eq:master}), to a theory 
that relies on interactions between particles such as that of quantum 
mechanics and the probabilistic interpretation of wave functions 
$|\psi ({\bf r})|^{2}$ \cite{Cur06}. 
Our hope is to stimulate further investigations in this direction
despite such possibility may not have at present convincing physical
ramifications and this paper may look rather metaphysical.

\section*{Acknowledgements}
Sincere thanks are due to Carlo Fonda for the graphics and an anonymous
referee for useful comments, questions and suggestions.

\begin{figure}[!t]
\includegraphics[width=15.7cm]{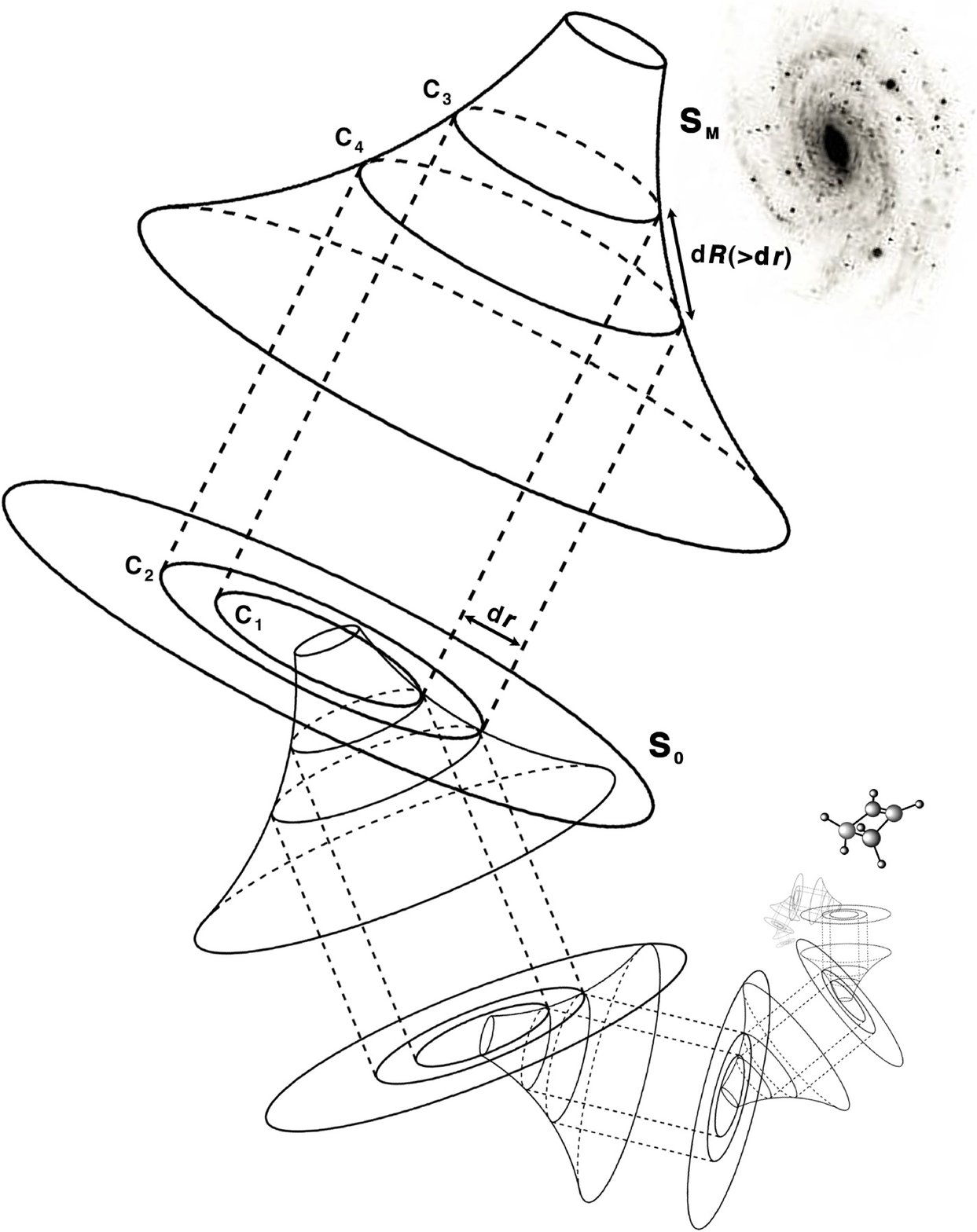}%
\caption{System of nested curved surfaces 
forming a spiral based on the Schwarzschild geometry 
due to the presence of dispersed mass
({\it c.f.}, Eq.(\ref{eq:dr1})) along distance scales from the macro- 
to the micro-world. 
}
\end{figure}

\end{document}